\documentclass[conference]{IEEEtran}
\usepackage{cite}

\usepackage{booktabs}
\usepackage{array}

\ifCLASSINFOpdf
\usepackage[pdftex]{graphicx}
\else
\fi
\usepackage[cmex10]{amsmath}
\usepackage{tabulary}

\usepackage{algorithmic}

\usepackage{mdwmath}
\usepackage{mdwtab}

\usepackage[tight,footnotesize]{subfigure}

\usepackage{placeins}

\begin{document}
%
\title{Single Stage PFC Flyback AC-DC Converter Design}  
\author{\IEEEauthorblockN{Kali Naraharisetti,  \textit{Member, IEEE,} Janamejaya Channegowda,  \textit{Member, IEEE,}}
}  


%


\maketitle

\begin{abstract} 
This paper discusses a 100 W single stage Power
Factor Correction (PFC) flyback converter operating in boundary mode constant
ON time methodology using a synchronous MOSFET rectifier on the
secondary side  to achieve higher efficiency. Unlike
conventional designs which use two stage approach such as PFC
plus a LLC resonant stage or a two stage PFC plus flyback, the proposed
design integrates the PFC and constant voltage
regulation in a single stage without compromising the efficiency of the converter. The proposed design is advantageous as it has a lower component count. A design of 100 W flyback operating from
universal input AC line voltage is demonstrated in this paper. The
experimental results show that the power factor (PF) is greater
than 0.92 and total harmonic distortion (iTHD) is less than 20\%
for a load varying from 25 \% to 100 \%. The experimental results
show the advantages of a single stage design.

\hspace*{-0.6cm} Keywords - Critical Conduction Mode (CrCM), Power Factor
Correction (PFC), Total harmonic distortion (iTHD), Flyback converter

\end{abstract}

\IEEEpeerreviewmaketitle

 \section{Introduction} 
All power supplies in the market with a power rating greater than 75 W
have to comply with an international regulation standard IEC
61000-3-2 \cite{1}. This standard puts a limitation on the current
harmonic percentage for class D power supplies in the power
range of 75 W to 600 W \cite{2},\cite{3}. Power supplies and power
adapters less than 75 W do not have a requirement on the total
harmonic distortion (iTHD) or power factor correction (PFC).

A power supply which does not have a PFC front end stage
draws the input current only when the diode rectified voltage is
greater than the DC link capacitor voltage \cite{4}. The pulsating
input current will have higher amounts of harmonics which
translates to power loss within the converter. In
other words, the power converter is drawing current more than
required. In low power LED applications (less than 50 W), a
single stage boundary mode PFC flyback converter is
commonly used because of low cost, wide input line and output
load variations and primary to secondary isolation capability \cite{5}-\cite{8}.
The power factor correction is done in a single stage approach
as well as a two stage approach.
\begin{itemize}
\item A single stage approach can use a flyback converter which can perform power factor correction and also voltage regulation

\item A two stage approach would use a PFC Boost converter followed by the flyback converter
\end{itemize}   

In a two stage approach a PFC boost converter would take the universal input
(85 $V_{ac}$ - 265 $V_{ac}$) and boost this input to a constant 400 V DC.
This 400 V DC is stepped down by second stage which can be a
flyback or an LLC resonant converter. Secondary stage flyback is
simple to design and leads to a lower bill of materials (BOM),
cost and lower efficiency compared to LLC resonant converter.
LLC resonant converter is a resonant topology which can give
higher efficiency compared to a hard switched topology by
providing Zero Voltage Switching (ZVS) to the primary half
bridge MOSFET switches. It also provides zero voltage and Zero
Current Switching (ZCS) to the secondary side synchronous rectifier MOSFETs. The downside of LLC resonant converter is
its complex design methodology, higher component count and
bill of materials cost.

\begin{figure}[!hbt] 
\centering
		\includegraphics[scale=0.47]{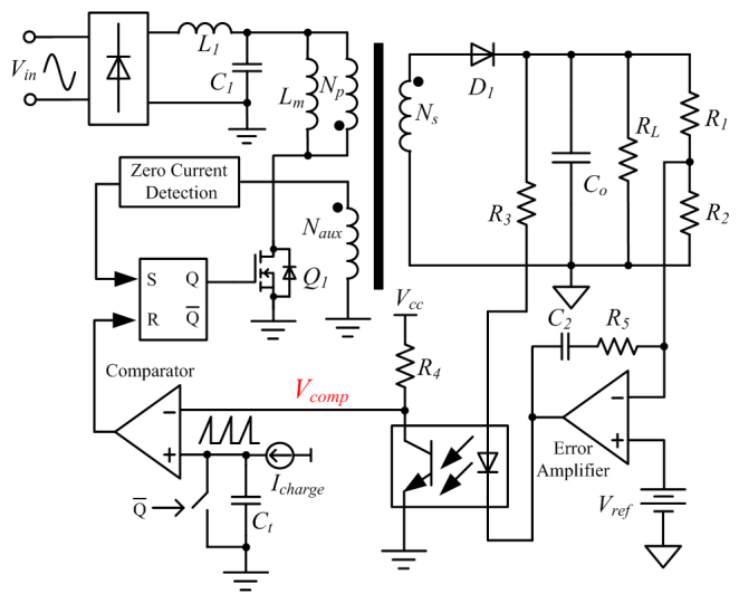}
	\caption{ Boundary mode flyback converter [5]}
		\label{fig:fig1}
		\end{figure}	
		
\begin{figure}[!hbt] 
\centering
		\includegraphics[scale=0.44]{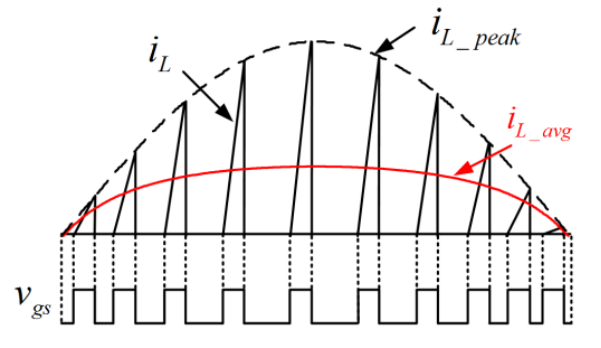}
	\caption{Theoretical transformer peak and average currents [5]}
		\label{fig:fig2}
		\end{figure}	
In \cite{9}, the authors use a complicated sliding mode controller to
implement PFC flyback with the use of extra
switches, but do not show a complete design of the converter.
Flyback implementation using discontinuous
conduction mode (DCM) is shown in \cite{8}. The problem with DCM mode is the
peak and the RMS currents are higher compared to critical
conduction mode or continuous conduction mode. The
implementation of the circuit in \cite{8} tends to be open loop and
the efficiency is low due to the losses in
the primary MOSFET and secondary side diode. 

In \cite{4}, the authors
implemented an extra gate drive circuit for the secondary side
synchronous MOSFET which increases the component count and
the complexity of the gate drive circuitry. In addition to this,
the power factor goes below 0.8 at the line voltage of 150 $V_{ac}$ -
230 $V_{ac}$ and the power rating is only 75 W.  In \cite{10},
the authors implement a variable ON time technique for an input AC cycle
by introducing a complex multiplier circuit at the output of an
optocoupler which needs extra gate pulses for the MOSFETs. The
results show crossover distortion at the input AC cycle zero crossing which increased the total harmonic distortion. 

The power levels of the converter in \cite{11} shows a complicated two stage
approach which uses a forward and flyback to achieve power
factor correction. This method shows higher voltage stress on
forward MOSFET which is approximately two times the input
voltage plus the extra leakage spikes, this has lead to loss in efficiency. In \cite{15} and \cite{17}, the authors discuss about primary side regulated flyback which becomes difficult to implement at 100 W,
especially if the regulation is through auxiliary winding. In
\cite{16}, authors use LLC half bridge resonant converter which is
suitable for 100 W design, but the disadvantages include higher BOM cost and increased
complexity.

In this paper, a 100 W single stage PFC
flyback is introduced. In Section II, the complete operation of
the flyback is explained. Section III briefly reviews the major
equations required to design the 100 W single stage flyback converter
running from a universal AC voltage of 85 $V_{ac}$ - 265 $V_{ac}$. Section
IV shows the results and waveforms of the complete design.

\section{Constant Voltage PFC Flyback} 
\label{sec:csv}

\subsection{Operation of the Power Factor Correction (PFC) flyback:} 

In Fig. \ref{fig:fig1} the basic diagram of a single stage flyback converter is shown.
When the primary MOSFET turns ON, the current in the
transformer primary ramps up as shown in Fig. \ref{fig:fig2}. During the
primary MOSFET ON time, the voltage across the primary
magnetizing inductance is equal to the input voltage. When
primary MOSFET is ON, the secondary side diode or synchronous
MOSFET remains OFF due to the direction of dot winding. 

The secondary diode sees a voltage of -$NV_{in}$ at the anode and at the cathode a voltage $V_{out}$ where N is the turns ratio. Since the current does not
circulate on the primary and secondary at the same time, the
flyback transformer is called a coupled inductor \cite{12}. When the
flyback PWM IC turns the primary MOSFET off, the voltage
across the primary inductor reverses in an attempt to keep the
ampere turns constant. The voltage across the primary MOSFET
is the sum of input voltage $V_{in}$ and the voltage across the
primary inductance $V_{Lp}$ .The secondary side diode has a positive voltage developed across it and starts conducting. The
energy stored in the inductance is transferred to the secondary
side thereby charging the output capacitor. 

During the duration when 
secondary diode is ON, the output voltage flies back to the
 primary and becomes $\frac{V_{out}}{N}$ \cite{12}. Hence the name flyback. In boundary mode flyback, the PWM IC IRS2982 turns ON the
primary MOSFET for a constant amount of time. Therefore the
current ramps up during ON time and ramps down during OFF
time based on the voltage applied across the primary inductor during ON time which is $V_{in}$ , and during OFF time - $\frac{V_{OUT}}{N}$. In this way, the average input current automatically follows the input rectified AC line voltage as shown in Fig. \ref{fig:fig2}. 

IRS2982 PWM IC eliminates the usage of conventional extra multiplier
circuit used from the DC bus since it uses the concept of
constant ON time.The OFF time is detected using an extra
auxiliary winding used in the transformer which detects the
zero crossing of the inductor current and gives a command to turn ON the primary MOSFET thereby starting a new switching
cycle \cite{1}, \cite{12}. The frequency in boundary mode is not fixed
and it is variable. The ON time is fixed and the off time is
variable. At low input AC voltage, the operating frequency is lower since
the duty ratio is higher whereas at higher input voltage, the
duty ratio is lower and the operating frequency is higher.

Flyback converter operating in boundary mode or critical
conduction mode has several advantages such as the power
stage of the complete converter appears as a first order system,
the reverse recovery of the secondary side diode does not pose problems and the primary inductance value is low . The secondary side diode need not be
a fast recovery diode. The limit on using boundary mode is
roughly around 200 W -250 W. The reason being the peak and
RMS currents increase as the power levels go up which
translate to higher power losses in the MOSFET and higher losses
in the transformer due to higher AC flux swing \cite{13}.

In order to regulate the output voltage, the regulation occurs
via a low-bandwidth closed-loop system, which in voltage
mode control maintains a constant on time to ensure a unity
power factor operation. In current mode, the controller imposes a
sinusoidal peak inductor current, without actively tracking the
average value of this current \cite{13}.

\subsection{Synchronous MOSFET rectifier:}

In order to improve the efficiency of the flyback converter, on the
secondary a synchronous MOSFET rectifier is used. When the
secondary side conducts, the current starts flowing through the
body diode of the secondary MOSFET, the voltage across the
MOSFET is sensed by the IR1161L IC and it turns ON the MOSFET,
thereby reducing the losses significantly from $V_f$ $\cdot$ $I_{OUT}$ to
 $R_{dson}$ $\cdot$ $I^2_{OUT}$.

\section{Flyback Converter Design}

In this paper a 100W PFC flyback design, on the secondary side a
TL431 is used which has an internal 2.5 V reference (Fig. \ref{fig:fig11}).
When the output voltage goes higher than 2.5V, the TL431
starts sinking current, thereby modulating the current in the
optocoupler diode which turns ON the internal BJT transistor.
The collector of this BJT transistor is connected to the COMP
pin of the IC IRS2982. Once the BJT turns ON, the COMP is
pulled low, thereby limiting the duty cycle. The COMP pin
decides the ON time of the primary MOSFET. If the COMP
voltage is high then the duty cycle will be longer (primary
MOSFET is ON for a longer duration) which translates to more energy being
transferred to the secondary side and vice versa. In Table I,
different parameters have been shown related to the 100 W PFC
flyback board design.

\begin{table}[htbp] 
\center
\caption{Parameters of the PFC Flyback}

\begin{tabular}{>{\flushleft}m{1.9in}     >{\flushleft}m{1in}}
\hline
 
\rule{0pt}{3ex}   \textbf{Parameters}  & \rule{0pt}{3ex}  \textbf{Value}   \\ \hline
\rule{0pt}{3ex} AC Input Voltage Range $(V_{ac})$ & \rule{0pt}{3ex} 85 - 264 V \\ \hline

\rule{0pt}{3ex} Output Voltage ($V_o$) & \rule{0pt}{3ex}  24 V  \\ \hline

\rule{0pt}{3ex} Output Current ($I_o$) & \rule{0pt}{3ex}  4.2 A \\ \hline

\rule{0pt}{3ex} Input Line Frequency ($Hz$)    &  60 Hz  \rule{0pt}{3ex}    \\ \hline

\rule{0pt}{3ex} Minimum Frequency ($f_{MIN}$)    & 70 kHz \rule{0pt}{3ex}    \\ \hline

\rule{0pt}{3ex}  Transformer Core   & EER35  \rule{0pt}{3ex}    \\ \hline

\rule{0pt}{3ex} Primary Inductance ($L_P$)    & 160 $\mu$H \rule{0pt}{3ex}    \\ \hline

\rule{0pt}{3ex} Primary Mosfet ($M_2$)  & IPD80R360P7S  \rule{0pt}{3ex}    \\ \hline

\rule{0pt}{3ex} Secondary Synchronous Mosfet  ($M_1$)  & IPB073N15N5  \rule{0pt}{3ex}    \\ \hline

\rule{0pt}{3ex} Primary PWM Controller ($IC_1$ )   & IRS2982S  \rule{0pt}{3ex}    \\ \hline

\rule{0pt}{3ex} Secondary Synchronous Controller ($U_1$)  & IR1161L  \rule{0pt}{3ex}    \\ \hline

\rule{0pt}{3ex} Output Capacitor ($C_o$)    &  1410 $\mu$F \rule{0pt}{3ex}   \\ \hline

\rule{0pt}{3ex} Maximum Duty Cycle ($D_{max}$ )
   & 0.5 \rule{0pt}{3ex}    \\ \hline

\rule{0pt}{3ex} Turns ($N_p:N_s:N_{aux}:N_{s_2}$)   & 40:6:5:3 \rule{0pt}{3ex}    \\ \hline

\rule{0pt}{3ex} Minimum Input Voltage $(V_{ac_{min}})$   & 85 V\rule{0pt}{3ex}    \\ \hline

\rule{0pt}{3ex} Output ripple $(V_{ripple})$   & 2 V \rule{0pt}{3ex}    \\ \hline

\rule{0pt}{3ex} Leakage Inductance
   & 3.75 $\mu$H \rule{0pt}{3ex}    \\ \hline

\rule{0pt}{3ex} Turns ratio $(n=\frac{N_s}{N_p})$   & 0.15 \rule{0pt}{3ex}    \\ \hline

\rule{0pt}{3ex} Peak Primary Current ($I_{peak}$)
   & 4.54 \rule{0pt}{3ex}    \\ \hline

\end{tabular}

\label{tab:para_flyback}
\end{table}
 
\textbf{Design Equations:}

Transformer inductance is determined (1) \cite{4},\cite{17}:
\begin{align}
L_{pri} = \frac{V^2_{acmin} \eta D^2_{max}}{\sqrt{2}P_{outmax}f_{min}}
\end{align}
 
\begin{itemize}
\item $L_{pri}$ is the primary inductance
\item $V_{acmin}$ is the minimum AC voltage
\item $D_{max}$ is the maximum duty cycle
\item $P_{outmax}$ is the maximum output power
\item $f_{min}$ is the minimum frequency
\end{itemize}
 
Turns ratio of the transformer is given by (2)

\begin{align}
\frac{N_S}{N_P} = \frac{V_{out}+V_f}{\sqrt{2}V_{acmin}} \cdot \frac{1-D_{max}}{D_{max}}
\end{align}

\begin{itemize}
\item $N_s$ is the secondary winding turns
\item $N_p$ is the primary number of turns
\item $V_f$ is the forward drop of the diode
\end{itemize}

Auxiliary to primary turns ratio is given by (3) and auxiliary to secondary turns ratio is given by (4)

\begin{align}
\frac{N_A}{N_P} &= \frac{V_{auxmax}+V_f}{\sqrt{2}V_{acmin}} \cdot \frac{1-D_{max}}{D_{max}} \\
\frac{N_A}{N_S} &= \frac{V_{auxmax}+V_f}{V_{out}+V_f}
\end{align}

\begin{itemize}
\item $N_A$ is the number of auxiliary winding
\end{itemize}
In a single stage flyback, the 120 Hz ripple is quite large compared to a two-stage topology. The higher the number of output capacitors, the lower is the output low frequency 120 Hz ripple. The output capacitance is calculated using the below formula: 
\begin{align}
C_{O}=\frac{I_{\text {out}}}{2 \cdot \pi \cdot f_{\text {acmin}} \cdot V_{\text {ripple}}}
\end{align}

\begin{itemize}
\item $C_o$ is the output capacitance
\item $I_{out}$  is the output current,
\item $f_{acmin}$ is  the minimum input AC line frequency
\item $V_{ripple}$ is the acceptable ripple on the output voltage
\end{itemize} 

 From the schematic
shown in Fig. \ref{fig:fig11}, the output capacitance used is around
1410 $\mu$F. The higher the capacitance, the lower is the 120 Hz
ripple. The leakage in a flyback transformer can be an issue for
the operation of the primary MOSFET. When the primary MOSFET in the converter
turns OFF, the leakage inductance of the transformer rings with
the $C_{oss}$ of the MOSFET. The leakage spike plus the input voltage
appears across the MOSFET. If the leakage spike is not clamped
or limited to a safe voltage lower than the breakdown voltage
of the MOSFET, then the MOSFET could easily be destroyed. In
order to limit the leakage spike, an RCD snubber is used \cite{14}.
The equations of the RC snubber are shown in (6).
\begin{align}
R_{S}=\frac{V_{s n}^{2}}{\frac{1}{2} \cdot L_{l k} \cdot I_{p e a k}^{2}\left(\frac{V_{s n}}{V_{s n}-n \cdot V_{0}}\right) f_{S}}
\end{align}
\begin{itemize}
\item $V_{sn}$  is the clamp voltage
\item $L_{lk}$ is the leakage inductance
\item n is the turns ratio
\item $I_{peak}$ is the peak primary transformer current
\end{itemize}

Snubber capacitor is given by (7)
\begin{align}
C_{S n}=\frac{V_{s n}}{\Delta V_{s n} R_{s n} f_{s}}
\end{align}

\begin{itemize}
\item $C_{sn}$ is the snubber capacitor 
\item $R_{sn}$  is the snubber resistor
\end{itemize} 
 
From Fig. 11  the snubber capacitor is a 2.2 nF/ 630 V and snubber resistor is a 2 W/ 75 k$\Omega$

 \begin{table}[htb!]
  \begin{center}
    \caption{Results at 120 Vac}
    \label{tab:table1}
    \begin{tabular}{c|c|c|c|c|c|c} 
      \textbf{$P_{IN}$} & \textbf{$V_{OUT}$} & \textbf{$I_{OUT}$} & $P_{OUT}$ & PF & $I_{THD}$ & $\eta$\\
(W) & (V) & (A) & (W) & & (\%) & (\%) \\       
\hline 114.68 & 23.96 & 4.20 & 100.8 & 0.99 & 6 & 87.90 \\
\hline 109.07 & 23.96 & 4.00 & 95.98 & 0.99 & 6.4 & 88.00 \\
\hline 102.09 & 23.96 & 3.75 & 89.99 & 0.99 & 4.9 & 88.15 \\
\hline 95.18 & 23.96 & 3.50 & 84.00 & 0.99 & 6.35 & 88.26 \\
\hline 88.15 & 23.96 & 3.25 & 78.01 & 0.99 & 4.45 & 88.50 \\
\hline 81.4 & 23.96 & 3.00 & 72.05 & 0.99 & 6.26 & 88.51 \\
\hline 74.63 & 23.96 & 2.75 & 66.03 & 0.99 & 6.2 & 88.48 \\
\hline 67.92 & 23.96 & 2.50 & 60.04 & 0.99 & 4 & 88.40 \\
\hline 54.52 & 23.96 & 2.00 & 48.06 & 0.99 & 7.95 & 88.16 \\
\hline 41.65 & 23.96 & 1.50 & 36.08 & 0.99 & 9.61 & 86.64 \\
\hline 28.86 & 23.97 & 1.00 & 24.09 & 0.99 & 10.53 & 83.47 \\
\hline 16.54 & 23.97 & 0.51 & 12.22 & 0.94 & 27.23 & 73.87 \\
\hline 9.26 & 23.97 & 0.27 & 6.52 & 0.90 & 32.46 & 67.74 \\
    \end{tabular}
  \end{center}
\end{table}

\section{Experimental Results}

Table II lists the power factor, THD, efficiency, and output voltage regulation at various loads starting from no load  to a full load of 100 W at 120 V input AC line voltage. As it can be seen from the table, the output voltage regulation over the load is constant. It can also be observed that the power factor is greater than 0.9 and THD is less than 11 \% from 25\% load to 100\% load.

Table III shows the results at 230 V AC input line voltage. The output voltage regulation is stable. Power factor is greater than 0.9  and THD is less than 20 \% from 25\% load to 100\% load. The efficiency of flyback converter is well within acceptable limits.

Figs. 3 and 4 show that the complete 100 W design met the
IEC61000 standard. Figure 5 shows the steady state waveforms of
converter at 120 Vac and figure 7 show the waveforms at 230 Vac.
The AC ripple on the output voltage can be reduced by adding
extra capacitors. Fig. 6 shows the start-up of the converter. As it
can be seen there is no overshoot on the output voltage at the
start up. The Figs. 8 and 9 show the steady state waveforms of the designed converter. Figs. 11 and 12 show the complete schematic of the flyback converter. From figures 13 and 14, it is clear that the flyback converter is very stable as phase margin is 40 degrees at 120 VAC and 54 degrees at 230 VAC.

Figure 10 illustrates the $V_{DS}$ and $V_{GS}$ waveforms of secondary side MOSFET at 230 VAC. When drain current on the secondary flows through the body diode of the secondary  MOSFET, the synchronous IC IR1161L uses the $V_{DS}$ sensing method to detect the negative $V_{DS}$ ( greater than -0.23 V) and turns ON the MOSFET gate to reduce the power losses in the diode. When the secondary current reaches close to zero, the IR1161L detects the $V_{DS}$  voltage and when it is in the range -8 mV to -4 mV , it  turns off the MOSFET gate. 
In this design, instead of using a simple diode on the secondary side, the design uses a MOSFET in  order to reduce the power losses in the diode from $V_f$ $\cdot$ $I_{out}$ to $I_{ds}^2$ $\cdot$ $R_{dson}$. By doing this, the heat sink requirement on the secondary side is eliminated by making use of a synchronous MOSFET.

 \begin{table}[htb!]
  \begin{center}
    \caption{Results at 230 Vac}
    \label{tab:table1}
    \begin{tabular}{c|c|c|c|c|c|c} 
      \textbf{$P_{IN}$} & \textbf{$V_{OUT}$} & \textbf{$I_{OUT}$} & $P_{OUT}$ & PF & $I_{THD}$ & $\eta$\\
(W) & (V) & (A) & (W) & & (\%) & (\%) \\       
 
\hline 111.76 & 23.97 & 4.212 & 100.96 & 0.988 & 9.43 & 90.34 \\
\hline 106.58 & 23.97 & 4.011 & 96.14 & 0.987 & 10.18 & 90.21 \\
\hline 100.11 & 23.97 & 3.761 & 90.15 & 0.986 & 11.11 & 90.05 \\
\hline 93.63 & 23.97 & 3.514 & 84.23 & 0.984 & 11.41 & 89.96 \\
\hline 87.35 & 23.97 & 3.264 & 78.24 & 0.981 & 12.58 & 89.57 \\
\hline 80.97 & 23.97 & 3.013 & 72.22 & 0.978 & 13 & 89.20 \\
\hline 74.63 & 23.97 & 2.763 & 66.23 & 0.975 & 12.21 & 88.74 \\
\hline 68.41 & 23.97 & 2.515 & 60.28 & 0.971 & 13.11 & 88.12 \\
\hline 55.75 & 23.97 & 2.013 & 48.25 & 0.952 & 13.49 & 86.55 \\
\hline 43.27 & 23.97 & 1.518 & 36.39 & 0.935 & 15.88 & 84.09 \\
\hline 31.28 & 23.97 & 1.015 & 24.33 & 0.891 & 19.62 & 77.78 \\
\hline 16.69 & 23.97 & 0.509 & 12.20 & 0.749 & 24.9 & 73.08 \\
\hline 10.192 & 23.97 & 0.261 & 6.26 & 0.581 & 37.82 & 61.38 \\ 
    \end{tabular}
  \end{center}
\end{table}

\begin{figure}[!hbt] 
\centering
		\includegraphics[scale=0.57]{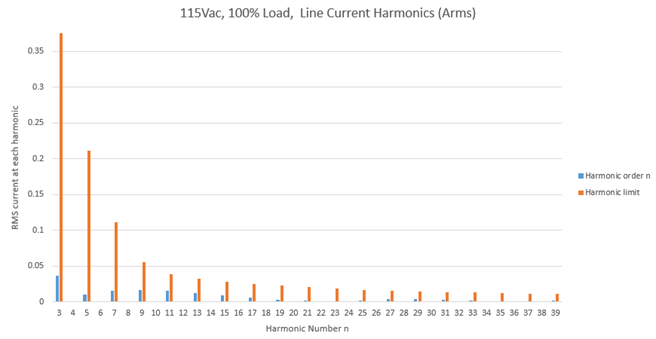}
	\caption{ IEC61000-3-2 Class D harmonic limits met at 120 Vac}
		\label{fig:fig3}
		\end{figure}

\begin{figure}[!hbt] 
\centering
		\includegraphics[scale=0.57]{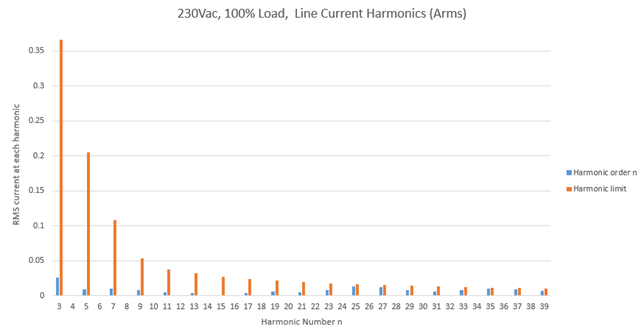}
	\caption{IEC61000-3-2 Class D harmonic limits met at 230 Vac}
		\label{fig:fig4}
		\end{figure}

\begin{figure}[!hbt] 
\centering
		\includegraphics[scale=0.5]{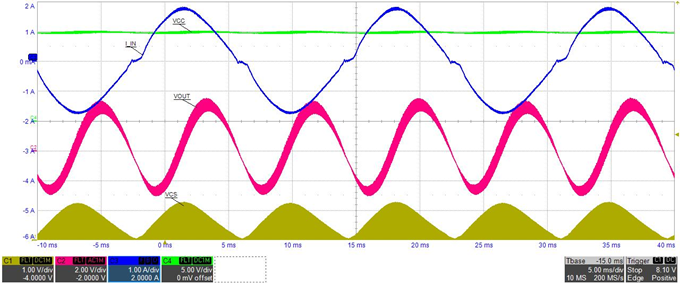}
	\caption{Steady state waveforms at 120 Vac: Input current (Blue), Current sense voltage (Yellow), Vout ripple (Red), Vcc (Green)}
		\label{fig:fig5}
		\end{figure}

\begin{figure}[!hbt] 
\centering
		\includegraphics[scale=0.5]{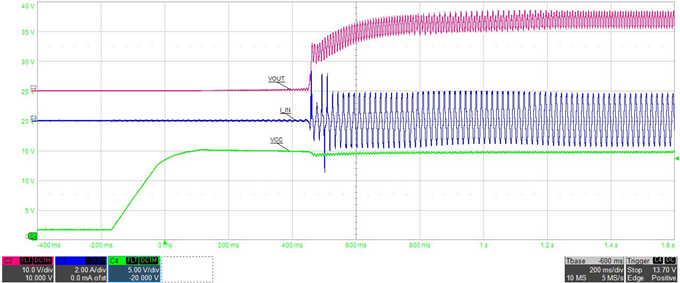}
	\caption{Startup at 120 Vac : Vout (Red), Input current (Blue), Vcc ( Green)
}
		\label{fig:fig6}
		\end{figure}

\begin{figure}[!hbt] 
\centering
		\includegraphics[scale=0.5]{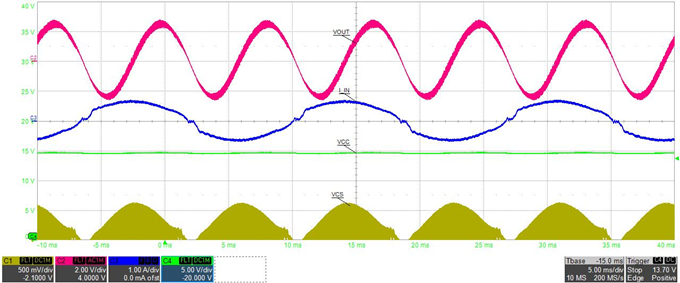}
	\caption{ Steady state waveforms at 230 Vac: Input current (Blue), Current sense
voltage (Yellow), Vout ripple (Red), Vcc (Green)
}
		\label{fig:fig7}
		\end{figure}

\begin{figure}[!hbt] 
\centering
		\includegraphics[scale=0.5]{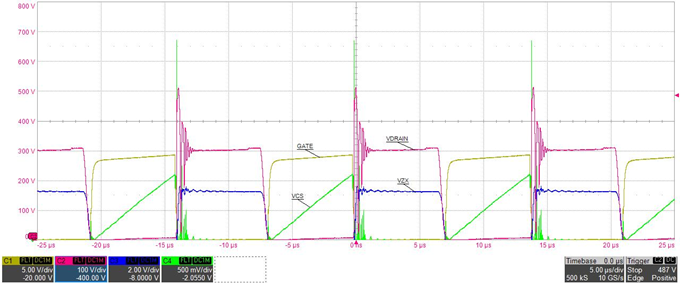}
	\caption{Steady state waveforms at 120 Vac: MOSFET drain voltage (Red), Current sense voltage (VCS:Green), MOSFET gate (Yellow), Zero crossing voltage at the auxiliary winding (VZX : Blue)}
		\label{fig:fig8}
		\end{figure}

\begin{figure}[!hbt] 
\centering
		\includegraphics[scale=0.5]{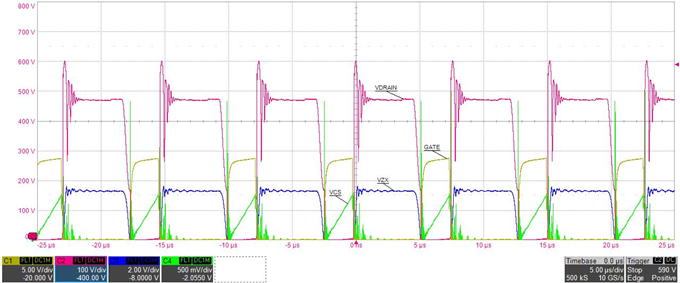}
	\caption{ Waveforms at 230 Vac: Drain voltage (Red), Current sense voltage (VCS: Green), Gate (Yellow), Zero crossing voltage (VZX : Blue)}
		\label{fig:fig9}
		\end{figure}

\begin{figure}[!hbt] 
\centering
		\includegraphics[scale=0.5]{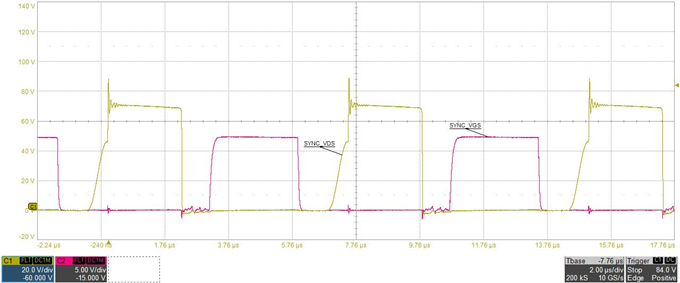}
	\caption{Synchronous rectification MOSFET waveforms at 230Vac, Drain voltage
of the synchronous MOSFET (Vds: Yellow), Gate of the MOSFET (Vgs: Red)
}
		\label{fig:fig10}
		\end{figure}

\begin{figure}[!hbt] 
\centering
		\includegraphics[scale=0.3]{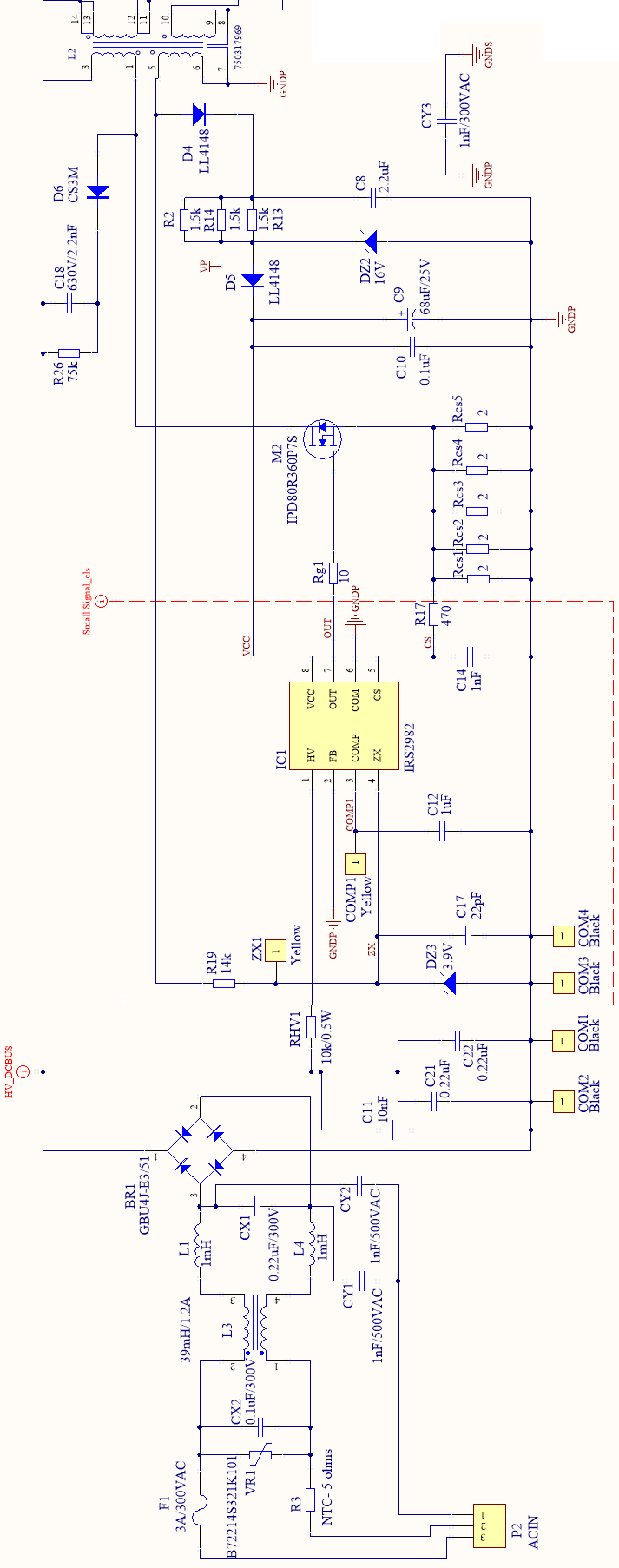}
	\caption{Complete schematic of 100 W single stage PFC Flyback}
		\label{fig:fig11}
		\end{figure}

\begin{figure}[!hbt] 
\centering
		\includegraphics[scale=0.32]{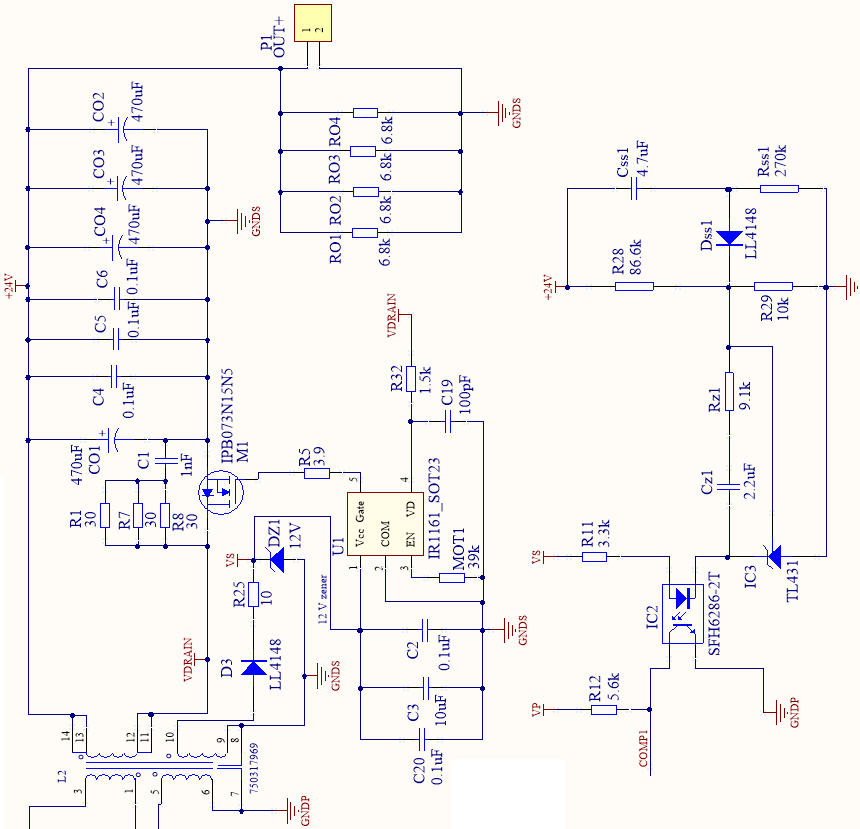}
	\caption{Secondary side of 100 W single stage PFC Flyback}
		\label{fig:fig12}
		\end{figure}
 
\begin{figure*}[!hbt] 
\centering
		\includegraphics[scale=0.43]{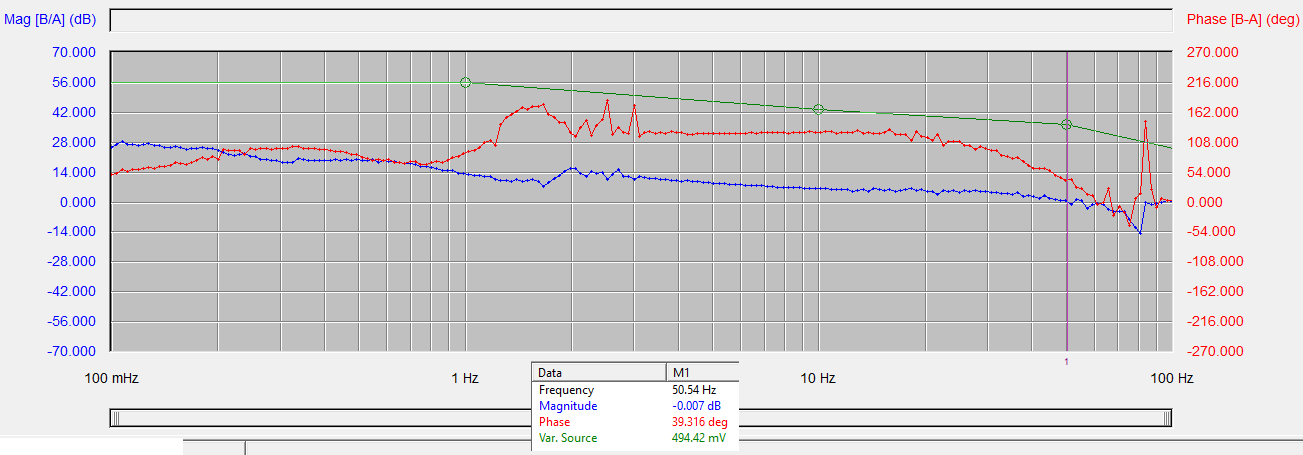}
	\caption{Phase Margin of the converter is 40 Deg at 120 VAC}
		\label{fig:fig13}
		\end{figure*}
 
\begin{figure*}[!hbt] 
\centering
		\includegraphics[scale=0.43]{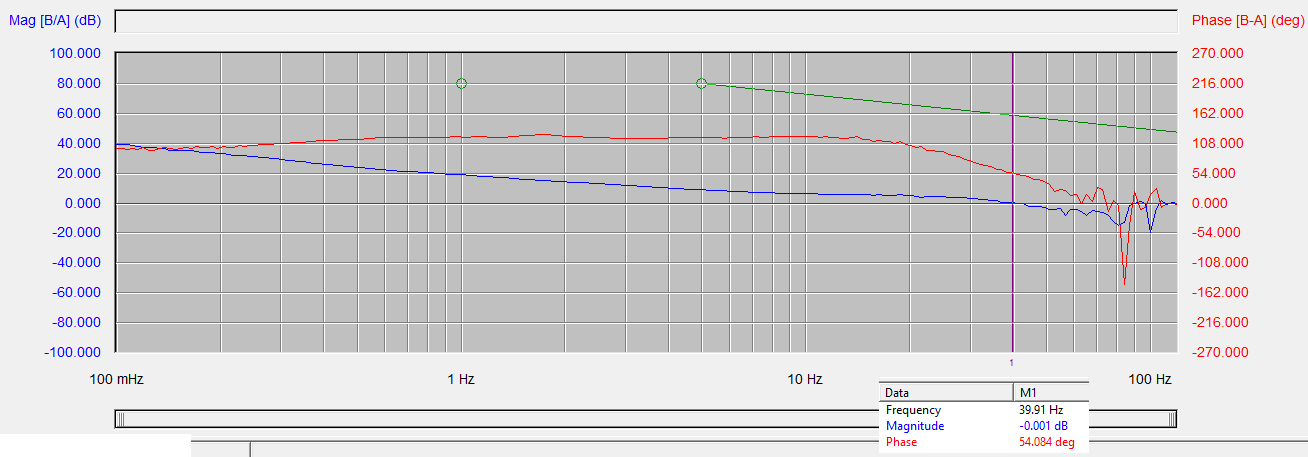}
	\caption{Phase Margin of the converter is 54 Deg at 230 VAC }
		\label{fig:fig14}
		\end{figure*}

\section{Conclusion}

In this paper a 100 W single stage PFC flyback design was presented. The results show that a single stage PFC flyback can be pushed to a higher power level thereby reducing the component count while maintaining a higher level of efficiency. The Power factor over the input voltage range was greater than 0.9 and THD remained less than 20\%.


\end{document}